\documentclass[fleqn,usenatbib,useAMS]{mnras}

\usepackage{graphicx}	
\usepackage{amsmath}	
\usepackage{amssymb}	
\usepackage{multicol}        
\usepackage{bm}		
\usepackage{pdflscape}	

\usepackage[T1]{fontenc}
\usepackage{ae,aecompl}

\usepackage{newtxtext,newtxmath}

\newcommand{\mdot}{\dot{M}}
\newcommand{\qdot}{\dot{q}}

\title[Magnetism, X-rays, and Accretion in WD\,1145]{Magnetism, X-rays, and Accretion Rates in WD\,1145+017 and other Polluted White Dwarf Systems}

\author[J. Farihi et al.]{J. Farihi$^{1}$\thanks{E-mail: j.farihi@ucl.ac.uk}\thanks{STFC Ernest Rutherford Fellow},
L. Fossati$^{2}$,
P. J. Wheatley$^{3}$,
B. D. Metzger$^{4}$,
J. Mauerhan$^{5}$,
S. Bachman$^{6,7}$,
\newauthor
B. T. G\"ansicke$^{3}$,
S. Redfield$^{7}$,
P. W. Cauley$^{7}$,
O. Kochukhov$^{8}$,
N. Achilleos$^{1}$,
N. Stone$^{9}$
\\
$^{1}$Department of Physics and Astronomy, University College London, London, UK\\
$^{2}$Space Research Institute, Austrian Academy of Sciences, Graz, Austria\\
$^{3}$Department of Physics, University of Warwick, Coventry, UK\\
$^{4}$Department of Physics, Columbia University, New York, USA\\
$^{5}$Department of Astronomy, University of California, Berkeley, USA\\
$^{6}$Department of Physics, Indiana University of Pennsylvania, Indiana, USA\\
$^{7}$Astronomy Department and Van Vleck Observatory, Wesleyan University, Middletown, USA\\
$^{8}$Department of Physics and Astronomy, Uppsala University, Uppsala, Sweden\\
$^{9}$Department of Astronomy, Columbia University, New York, USA\\
}

\begin{document}


\maketitle

\begin{abstract}

This paper reports circular spectropolarimetry and X-ray observations of several polluted white dwarfs including 
WD\,1145+017, with the aim to constrain the behavior of disk material and instantaneous accretion rates in these evolved 
planetary systems.  Two stars with previously observed Zeeman splitting, WD\,0322--019 and WD\,2105--820, are detected 
above $5\upsigma$ and $\langle B_{\rm z} \rangle > 1$\,kG, while WD\,1145+017, WD\,1929+011, and WD\,2326+049 yield 
(null) detections below this minimum level of confidence.  For these latter three stars, high-resolution spectra and atmospheric 
modeling are used to obtain limits on magnetic field strengths via the absence of Zeeman splitting, finding $B_* < 20$\,kG 
based on data with resolving power $R\approx40\,000$.  An analytical framework is presented for bulk Earth composition 
material falling onto the magnetic polar regions of white dwarfs, where X-rays and cyclotron radiation may contribute to 
accretion luminosity.  This analysis is applied to X-ray data for WD\,1145+017, WD\,1729+371, and WD\,2326+049, and 
the upper bound count rates are modeled with spectra for a range of plasma $kT = 1-10$\,keV in both the magnetic and 
non-magnetic accretion regimes.  The results for all three stars are consistent with a typical dusty white dwarf in a steady
state at $10^8 - 10^9$\,g\,s$^{-1}$.  In particular, the non-magnetic limits for WD\,1145+017 are found to be well below 
previous estimates of up to $10^{12}$\,g\,s$^{-1}$, and likely below $10^{10}$\,g\,s$^{-1}$, thus suggesting the star-disk 
system may be average in its evolutionary state, and only special in viewing geometry.

\end{abstract}


\section{Introduction}

It is now abundantly clear that a significant fraction of planetary systems survive the gauntlet of stellar evolution and manifest 
around white dwarf stars \citep{far16,ver16}.  These evolved planetary systems provide empirical and theoretical constraints 
on planet formation and evolution that are complimentary to conventional studies, and most importantly provide information
unavailable via pre-main and main-sequence stars.  Protoplanetary disks reveal ongoing chemical and spatial processes
related to the earliest stages of planet formation, and especially volatile grain chemistry and dust trapping \citep{pon14,van16} 
prior to and possibly during their incorporation into large bodies.  Young and mature main-sequence stars reveal systems of 
fully-fledged planets, their planetesimal belt leftovers, and (giant impact) collisional by-products \citep{men14,mac17,cha17}.
These planetary systems yield final architectures and assemblies, where the physical and chemical processes associated 
with planet formation are essentially exhausted and thus must be inferred via modeling.  White dwarfs are the only systems 
to provide bulk chemical information on large planetesimals and an alternative window onto planetary systems born around 
A and F-type stars.

More than a decade of observational evidence supports a picture where minor or major planets experience catastrophic
fragmentation within the Roche limit of white dwarfs \citep{jur03}.  This paradigm is supported by multiple lines of evidence
including atmospheric pollution via heavy elements \citep{zuc03,far10,koe14}, infrared and optical emission from closely
orbiting disks of dust and gas \citep{far16}, and chemical abundances broadly consistent with terrestrial-like planetesimals
\citep{gan12,jur14,wil16}.  Evidence for short-term disk evolution has been observed in gas and dust emission features, 
including evidence of eccentric disk precession \citep{xu14,wil14,man16}.  Arguably the most spectacular example of 
ongoing change in a white dwarf debris disk is the rapidly varying (over minutes to months) extinction observed in both 
photometry and spectroscopy towards WD\,1145+017 \citep{van15,gan16,rap16,red17}.

Theoretical progress on evolved planetary system dynamics, dust production, and disk evolution has been invaluable for 
placing the observations in context.  The lifetime of dust grains in flat disks that are (vertically) optically thick, will be driven 
by Poynting-Robertson drag \citep{raf11}, as this configuration can effectively damp grain-grain collisions that would otherwise 
be more than an order of magnitude more rapid \citep{far08}.  The presence of gas within disks may influence their evolution 
by dust-gas coupling \citep{met12} or $\alpha$-disk viscous dissipation \citep{jur08}, and this may result in shorter disk lifetimes 
than by dust evolution alone \citep{boc11}.  There is indirect evidence for historical changes in accretion rate over pollution 
lifetime seen in samples of stars with disparate metal sinking timescales \citep{gir12}, and this appears to arise from relatively 
short-lived episodes of dust and gas production \citep{far12,wya14}.  Recent work suggests that collisional cascades can 
dominate disk behavior at orbital radii comparable to the Roche limit, resulting in disks with significant vertical scale heights 
and substantially shorter lifetimes in the absence of replenishment \citep{ken17}.

The prevalent model for WD\,1145+017 involves at least one parent body orbiting near the Roche limit and losing mass in a
tail-like and possibly a head-like feature, potentially in discrete events that are eventually transformed into large clouds of gas
and debris \citep{van15,rap16}.  In this scenario the parent body or bodies are comparable in mass to the largest solar system 
asteroids, and this in turn places strong constraints on orbital and structural stability, implying essentially circular orbits and 
a relatively brief time period until total disintegration \citep{ver16b,ver17,gur17}.  Outstanding issues include the apparently 
circular orbit of one or more large planetary bodies at the Roche limit, and the likelihood of witnessing a short-lived event at 
a narrow range of possible viewing angles.  Recently, a novel model proposed that dust trapping in the stellar magnetosphere 
could, under favorable circumstances, be consistent with the quasi-periodic extinction seen towards WD\,1145+017 \citep{far17b}.  

Previous observation-based analyses suggested that WD\,1145+017 may be currently accreting at a rate significantly higher 
than inferred to be ongoing for any polluted white dwarf, and up to $10^{12}$\,g\,s$^{-1}$ \citep{xu16,gan16,rap16}.  This rate 
is $\sim10^2$ times higher than the steady-state inference from the heavy elements in the stellar atmosphere, and  $\sim10^3$ 
times higher than any instantaneous rate inferred via a steady-state regime calculation.

This paper presents X-ray observations to confirm or rule out the previously proposed high accretion rate for WD\,1145+017,
and corresponding circular spectropolarimetry to determine if either circumstellar dust, or the process of accretion is influenced
by stellar magnetism.  The observational data are presented in Section 2 with the resulting upper limits and physical constraints 
derived in Section 3.  This analysis includes a set of models that link magnetic field strength and accretion rate with the emergent
luminosity of the cooling flow and corresponding physical mechanisms, including X-ray emission.  Section 3 also includes a new 
high-resolution spectrum of WD\,1145+017 that indicates the observed circumstellar gas is not in the process of accreting.
Section 4 provides a summary of constraints for the accretion rates onto WD\,1145+017 and other polluted white dwarfs, as 
well as conclusions.

The remainder of the paper refers to stars by their numerical designation alone.

\section{Observations and Data}

As magnetic fields will influence the flow and luminosity of the accretion, the observational data appear in the following natural 
order.

\subsection{Optical Spectropolarimetry}

\begin{table*}
\begin{center}
\caption{FORS2 Spectropolarimetry and Longitudinal Magnetic Field Measurements\label{tbl1}}
\begin{tabular}{@{}ccccccccccc@{}}

\hline
\hline

WD\# 		&$V$	&SpT$^{\rm a}$	&$T_{\rm eff}$	&Date	&Grating		&Coverage	&$t_{\rm exp}$	&S/N		&$\langle B_{\rm z} \rangle^{\rm b}$	&$\langle N_{\rm z} \rangle$\\
 			&(mag)	&		&(K)			&		&		&(\AA)		&(s)			&					&(kG)						&(kG)\\

\hline

0322--019		&16.1	&DZA	&5300	&2016 Oct 04	&1200B+97	&$3660-5110$	&$4\times310$	&70		&$-5.4\pm3.0$ \ ($1.8\upsigma$)					&$+4.3\pm2.6$ \ ($1.6\upsigma$)\\ 
			&		&		&		&2016 Oct 05	&1200B+97	&$3660-5110$	&$4\times310$	&90		&$-$\,{\bf 16.5}\,$\pm$\,{\bf 2.3} \ ({\bf 7.1}$\upsigma$)	&$-1.7\pm2.1$ \ ($0.8\upsigma$)\\
1145+017		&17.2	&DBZA	&15\,900	&2017 Jan 01	&1200B+97	&$3660-5110$	&$8\times594$ 	&130		&$-0.1\pm0.4$ \ ($0.2\upsigma$)					&$-0.7\pm0.4$ \ ($1.7\upsigma$)\\ 
			&		&		&		&2017 Jan 02	&1200B+97	&$3660-5110$	&$4\times594$ 	&110		&$-1.0\pm0.4$ \ ($2.4\upsigma$)					&$-0.2\pm0.4$ \ ($0.4\upsigma$)\\ 
			&		&		&		&2017 Feb 01	&1200B+97	&$3660-5110$	&$4\times594$	&120		&$+$\,{\bf 1.0}\,$\pm$\,{\bf 0.3} \ ({\bf 3.5}$\upsigma$)	&$-0.1\pm0.3$ \ ($0.5\upsigma$)\\ 
			&		&		&		&2017 Feb 27	&1200B+97	&$3660-5110$	&$4\times594$ 	&110		&$-0.6\pm0.4$ \ ($1.5\upsigma$)					&$+0.5\pm0.4$ \ ($1.2\upsigma$)\\ 
1929+011		&14.2	&DAZ	&21\,200	&2016 Aug 04	&1200R+93	&$5750-7310$	&$4\times500$ 	&320		&$+0.5\pm0.4$ \ ($1.2\upsigma$)					&$+0.1\pm0.4$	\ ($0.3\upsigma$)\\ 
2105--820		&13.8	&DAZ	&10\,200	&2016 Jun 30	&1200R+93	&$5750-7310$	&$4\times300$ 	&350		&$+$\,{\bf 5.1}\,$\pm$\,{\bf 0.3} \ ({\bf 15.2}$\upsigma$)	&$-0.2\pm0.3$ \ ($0.6\upsigma$)\\ 
2326+049		&13.0	&DAZ	&11\,900	&2016 Aug 04	&1200R+93	&$5750-7310$	&$4\times120$	&290		&$-0.7\pm0.5$ \ ($1.4\upsigma$)					&$+0.1\pm0.5$ \ ($0.2\upsigma$)\\

\hline
\hline

\end{tabular}
\end{center}

\flushleft
$^{\rm a}$  White dwarf spectral types begin with `D' for degenerate star, followed by letters corresponding to elements in 
decreasing line strengths: e.g.\ `A' for Balmer absorption, `B' for He\,{\sc i} lines, and `Z' for metal features \citep{mcc99}.

\smallskip
$^{\rm b}$ P98 measurements are for metal and He\,{\sc i} lines, while P97 determinations are based on H$\alpha$ (see \S2.1).

\end{table*}

Five metal-rich white dwarfs -- including 1145+017 -- were observed as part of an ongoing search for weak stellar magnetism in 
evolved planetary systems where accretion from circumstellar matter is evident.  Each star was observed in spectropolarimetric 
mode with the Focal Reducer and low dispersion Spectrograph (FORS2; \citealt{app98}), which is mounted on UT1 of the ESO 
Very Large Telescope (VLT).  The targets were observed using either the 1200R grating and a $1\farcs0$ slit, or with the 1200B 
grating and a $0\farcs7$ slit, and all exposures were performed alternating the position of the quarter-wave retarder plate at 
$\pm45\degr$.  All data were taken in service mode with the MIT red-sensitive detector (the E2V blue-sensitive chip is only 
available in visitor mode), and read out using $2\times2$ binning.  Table \ref{tbl1} lists the target properties and observing 
details.

Three $T_{\rm eff} > 10\,000$\,K DAZ stars were observed over the H$\alpha$ region as this line is the strongest in their
optical spectra, and their weak metal lines are not readily detectable at the modest resolution of FORS2: 1929+011 (= 
GALEX\,J193156.8+011745), 2105--820 (= LTT\,8381), and 2326+049 (= G29-38).  These three stars were chosen in P97 to 
be relatively nearby and bright examples of polluted white dwarfs, and thus where the strongest constraints could be placed on 
stellar magnetism.  In P98, 0322--019 (= G77-50) and 1145+017 were observed as control and science  targets, respectively, 
where their distinct optical spectra required coverage at blue wavelengths.  For both stars the Ca\,{\sc ii} H and K lines are 
the deepest spectral features, with weak or undetectable H lines at modest resolution \citep{far11,xu16}.

The data were reduced and analyzed using a set of {\sc iraf} and {\sc idl} routines described in detail by \citet{fos15}, developed
based on previously published techniques and algorithms for spectropolarimetry \citep{bag12,bag13}.  Because of cosmic
rays affecting the data obtained in P98, the spectra were retrieved via weighted (optimal) extraction.

The surface-averaged, longitudinal magnetic field $\langle B_{\rm z} \rangle$ was measured using the following formula
\citep{ang70,lan75} and least-squares fitting \citep{bag02,bag12}:
\begin{equation}
V_\uplambda = -g_{\rm eff} \, C_{\rm z} \, \uplambda^2 \frac{1}{I_\uplambda} \frac{{\rm d}I_\uplambda}{{\rm d}\uplambda} \langle B_{\rm z} \rangle
\label{eqn1}
\end{equation}
where $V_\uplambda$ and $I_\uplambda$ are the Stokes $V$ and $I$ profiles, respectively.  The effective Land\'e factor, $g
_{\rm eff}$, was set to be 1.25 except in the region of the Balmer lines where it was set to 1.0.  The constant in the equation is 
defined as $C_{\rm z}=e/4 \uppi m_ec^2 \approx 4.7\,\times\,10^{-13}$\,\AA$^{-1}$\,G$^{-1}$, where $e$ is the electron charge, 
$m_{\rm e}$ the electron mass, and $c$ the speed of light.

In addition to $\langle B_{\rm z} \rangle$, $\langle N_{\rm z} \rangle$ was also measured and denotes the value of the 
surface-averaged, longitudinal magnetic field obtained from the diagnostic $N$ profile \citep{don97} in place of Stokes $V$. The 
$N$ profile is in practice the difference between an even number of Stokes $V$ profiles and as such provides a measure of the 
noise in the spectra, and $\langle N_{\rm z} \rangle$ can highlight possible spurious detections.  In addition, the $N$ profile 
permits statistical checks for an improved assessment of any genuine magnetic fields \citep{bag13,fos15}.  The Stokes $V$ 
and $N$ parameter spectra of 1145+017 showed rather  large-scale deviations from zero, and these spectra were therefore 
renormalized using high-order polynomials.

The $\langle B_{\rm z} \rangle$ and $\langle N_{\rm z} \rangle$ values were derived considering spectral windows covering 
either only hydrogen lines, or only metallic lines (including He\,{\sc i}).  Owing to the wavelength coverage, modest spectral 
resolution, and intrinsic weakness of their metal features, spectropolarimetry was performed only on the strong H$\alpha$ 
features of the P97 targets 1929+011, 2105--820, and 2326+049.  In contrast, the P98 targets 0322--019 and 1145+017 have 
hydrogen lines that are relatively weak compared to their other absorption features, and hence spectropolarimetry for these 
stars was carried out on metal and He\,{\sc i} lines.  A careful selection of lines was done for 1145+017 based on a HIRES 
spectrum with resolving power $R\approx40\,000$ as a reference.  Results are listed in Table \ref{tbl1}.

Magnetic field measurements conducted with low-resolution spectropolarimeters mounted at Cassegrain focus, such as 
FORS2, may be affected by strong systematics that may be difficult to characterize (see \citealt{bag12,bag13}).  In some 
cases, such systematics can lead to spurious $3\upsigma-4\upsigma$ detections, and for this reason a magnetic field is only
considered detected when above the 5$\upsigma$ level and with a $\langle N_{\rm z} \rangle$ value consistent with zero.  
However, for targets in which the measured $\langle B_{\rm z} \rangle$ value lies in the range $3\upsigma-5\upsigma$, and 
which also exhibit $\langle N_{\rm z} \rangle$ values consistent with zero, the general recommendation is to follow up with 
additional observations to assess the genuine presence of a structured large-scale magnetic field.

\begin{figure*}
\includegraphics[width=144mm]{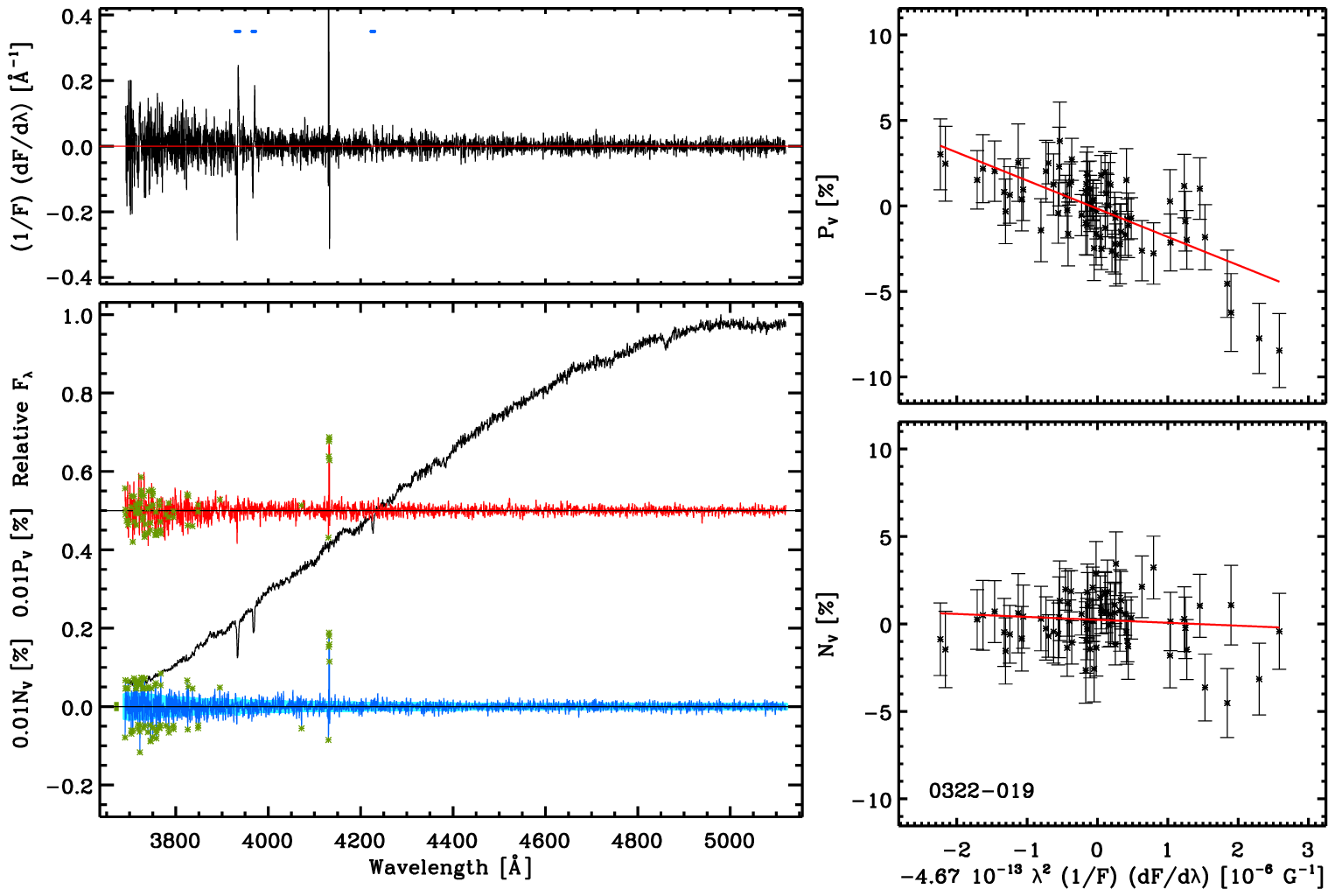}
\vskip 10pt
\includegraphics[width=144mm]{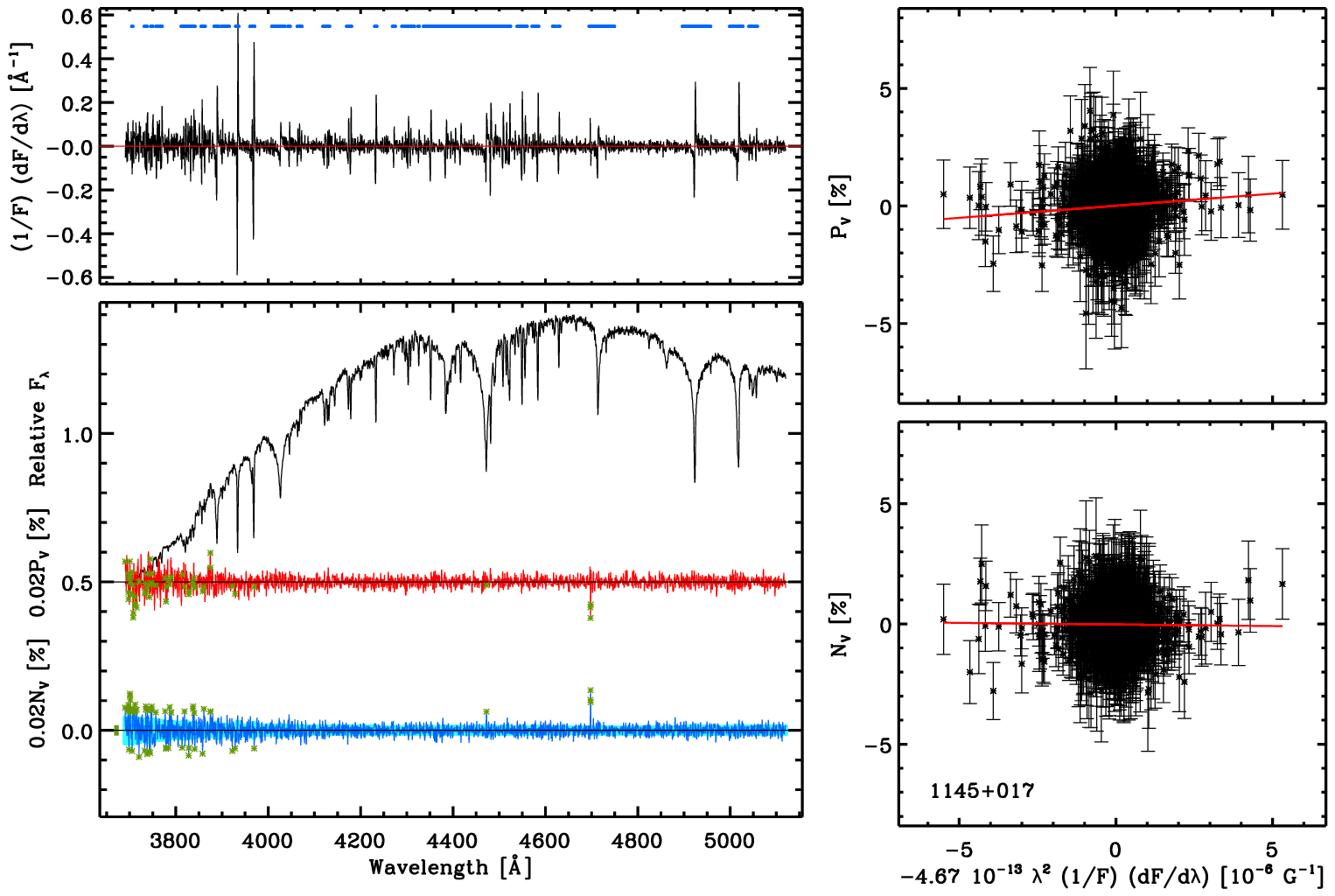}
\caption{Plots of spectropolarimetric data where the fractional polarization is greater than $3\upsigma-5\upsigma$ (see \S2.1
and boldface numbers in Table \ref{tbl1} column ten).  Each group of four plots corresponds to a single observational data set 
for one star.  The upper left panels plot the derivative of the slope for each spectrum in black, with blue dashes marking the 
regions used to determine $P_{\rm v}=V/I$.  The lower left panel shows the relative Stokes $I$ (total flux) spectrum in black, 
with $P_{\rm v}$ and $N_{\rm v}$ plotted against wavelength in red and blue respectively.  Green crosses label data points 
that were removed from analysis via sigma clipping of outliers from the local average (i.e.\ cosmic rays, detector artifacts, 
noisy regions).  The right-hand panels plot $P_{\rm v}$ and $N_{\rm v}$ as a function of the right-hand side of Equation 
\ref{eqn1}, where the red lines have been determined by least squares fitting, and the resulting slopes are exactly $\langle 
B_{\rm z} \rangle$ and $\langle N_{\rm z} \rangle$ respectively.
\label{fig1}}
\end{figure*}

\begin{figure*}
\centerline{\includegraphics[width=144mm]{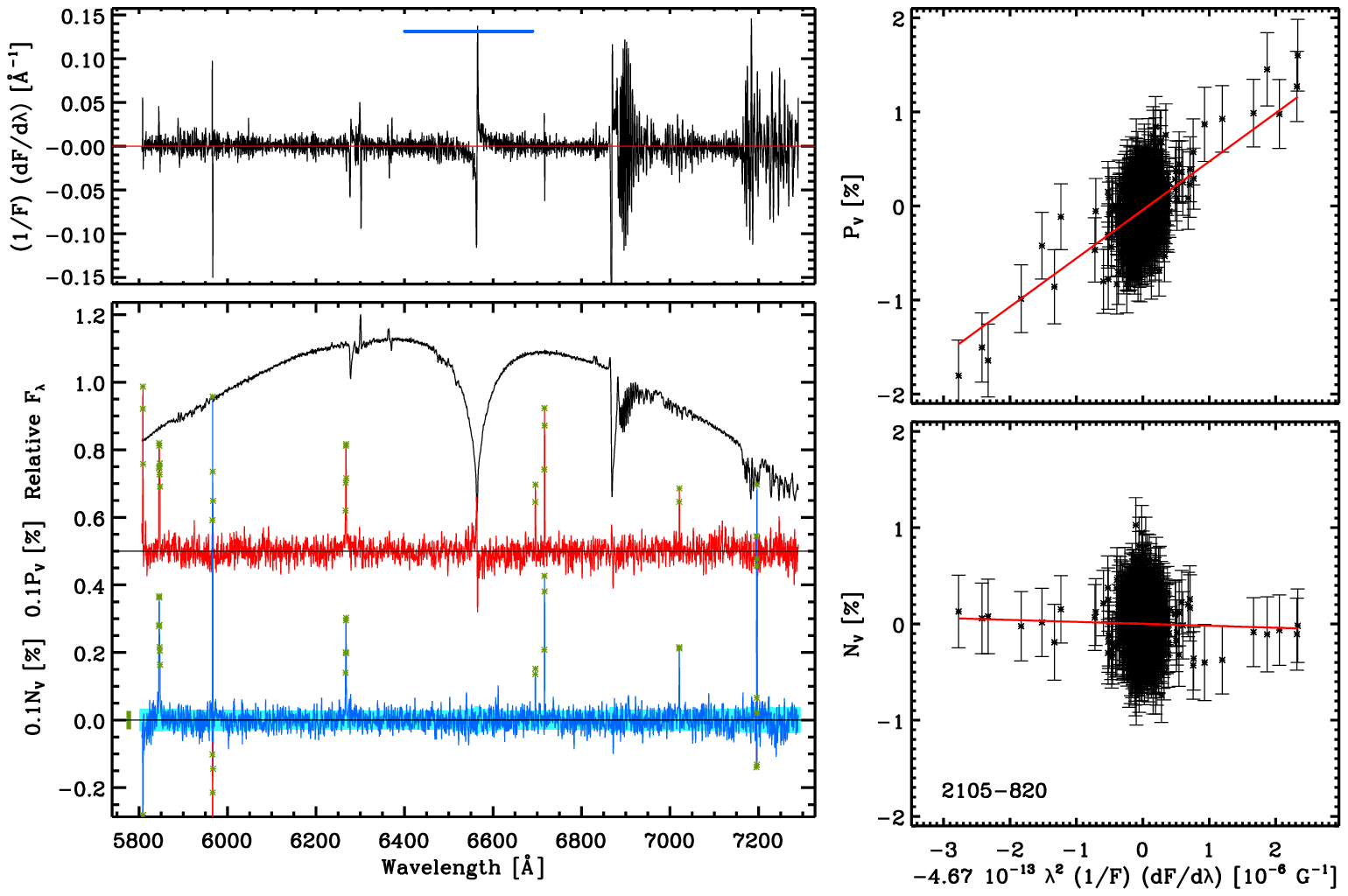}}
\contcaption{}
\end{figure*}

\subsection{X-rays}

\begin{figure*}
\includegraphics[height=168mm]{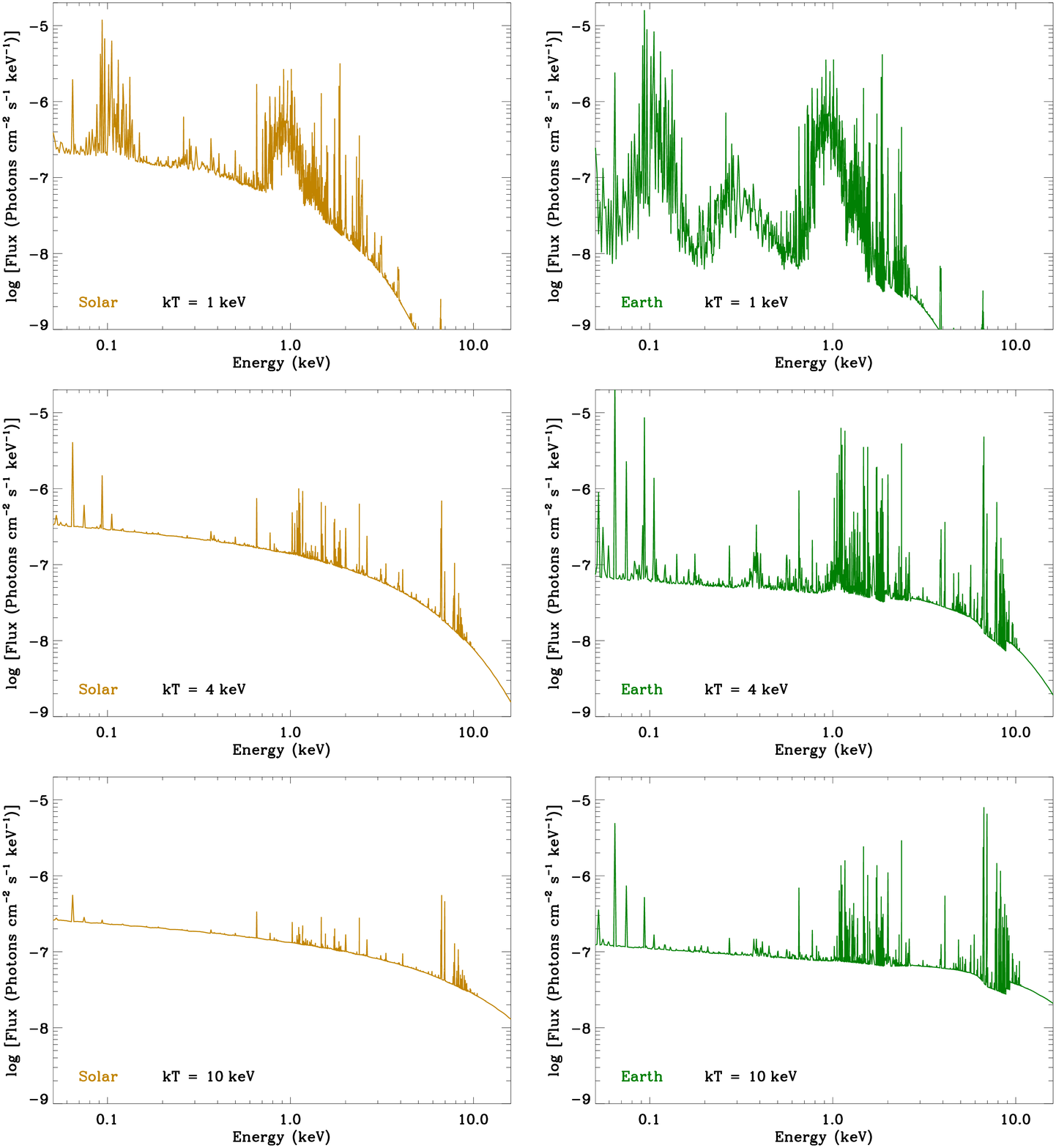}
\caption{Output of {\sc xspec} simulations run for solar (gold) and bulk Earth (green) composition plasmas of temperature 
1, 4, and 10\,keV.  Each of the plots has been normalized to match the MOS upper limit fluxes of WD\,1145+017.  The 
logarithmic scale emphasizes that the emergent flux is generally dominated by line emission, especially for lower temperatures 
and Earth composition gas.  These results support the analytical model (\S3.2) in which line emission -- from Fe in particular -- 
dominates the X-ray flux in the range $kT \simeq 1-10$\,keV.
\label{fig2}}
\end{figure*}

1145+017 was observed as a ToO with the {\em X-ray Multi-Mirror Mission (XMM-Newton)} beginning 2016 Jun 06 for a 
continuous duration of 134.9\,ks (observation \#0790181301).  Two additional polluted white dwarfs have previously been 
observed at X-ray wavelengths, both with {\em XMM}: 1729+371 (= GD\,362) and 2326+049 \citep{jur09}, and their data were
retrieved from the archive and re-analyzed here in a novel manner, as the detectability of X-rays from polluted white dwarfs in 
general is a focus of the present study.

Using the pipeline processed events list for 1145+017, there is no convincing evidence of a detection in either the pn or MOS
detectors.  The source position was corrected for the $\mu<0\farcs05$\,yr$^{-1}$ proper motion of the target, and counts were 
extracted from a 15\arcsec radius region centered at 11$^{\rm h}$\,48$^{\rm m}$\,33.58$^{\rm s}$\,+01\degr\,28\arcmin\,59.3\arcsec.  
The astrometric solution was also tested by inspecting other sources, and was found reliable to at least a few arcseconds.  Light 
curves were constructed to search for a possible detection in any short time interval, but none were apparent.  Data were filtered
for background flares caused by solar soft protons, when the $10-12$\,keV count rate from the whole camera was above 
0.5\,counts\,s$^{-1}$ for the pn, and above 0.2 and 0.3\,counts\,s$^{-1}$ for MOS1 and MOS2 detectors respectively.  This 
reduced the weighted on-source live time to 104.7\,ks for the pn detector, and 126.0\,ks for the MOS detectors (including the 
vignetting correction). 

The standard data reduction methods were followed, as described in data analysis threads provided with the Science Analysis 
System (SAS version 16.0).  A spectrum was extracted from the source position, and an adjacent 67\arcsec radius region free 
of sources was used to estimate the background.  Extractions were done for the energy ranges $0.3-2.0$ and $0.3-10$\,keV 
on the pn and MOS detectors, as well as the full range of the pn detector $0.16-12.0$\,keV.  Pipeline event files were used
throughout, but the data were independently reprocessed and no significant differences were found for events with pattern 
$\leq12$ for the MOS and pattern $\leq4$ for the pn detector (which are the standard filtering choices).  The data were 
checked with pattern $=0$ (single pixel) events in case this reduced the background more than the sensitivity, but this did 
not improve the count rate limits.  

Archival data for 1729+371 (20.4\,ks) and 2326+049 (22.5\,ks) were retrieved and similarly analyzed.  For all three targets, 
the background counts on the MOS detectors were found to be modestly to significantly lower than for the pn detector, thus 
yielding tighter constraints on source counts.  For this reason, the derived count rate constraints were taken from the combined 
MOS1 and MOS2 detector results for all sources.  Statistical confidence bounds on the number of source counts in each 
energy range were calculated following \citet{kra91}, which is a Bayesian approach designed for low counts that adopts the 
prior that source counts cannot be negative.  These 90\% confidence bounds on the source counts are not upper limits, strictly 
speaking, and are listed in Table \ref{tbl2}.  It is noteworthy that the upper bound MOS counts derived here for 2326+049 are 
approximately a factor of three higher than those previously published by \citet{jur09}, and almost certainly due to contamination 
from a point source $15''$ distant.  \citet{jur09} provide no details of corrective measures taken to remove or model the nearby
source contamination, and the counts and rates listed here for 2326+049 are unmodified.

The X-ray spectral fitting package {\sc xspec} 12.9 was used to calculate fluxes corresponding to the upper confidence bounds 
for given source spectra.  {\sc xspec} automatically accounts for the {\em XMM} instrument response and aperture correction at 
the target position, and allows for flexibility in model definition.  The range of adopted models assumed optically-thin plasmas, 
interstellar absorption from both grains and neutral hydrogen \citep{wil00}, with input abundances of solar \citep{asp09}, 
chondritic \citep{lod03}, and bulk Earth \citep{mcd00} compositions.  The results using the chondritic and bulk Earth 
compositions were found to be similar, and the former was discarded from the analysis. 

Interstellar absorption for 1145+017 was assumed to have $N_{\rm H} = 10^{20}$\,cm$^{-2}$, which is consistent with the 
observed hydrogen column density to WD\,1034+001 \citep{oli06}, a star in the same direction at a comparable distance.  This
adopted value is also consistent with the survey of hydrogen column densities by \citet{lin06}, where practically all objects within 
100\,pc of the Sun have $N_{\rm H} \leq 10^{19}$\,cm$^{-2}$, and all objects within 200\,pc have $N_{\rm H} \leq 10^{20}$\,cm$
^{-2}$.   The hydrogen column density toward 1729+371 and 2326+049 can be estimated in a similar way: for the former, using 
$\iota$\,Oph from \citet{val93}, a value of $N_{\rm H} \approx 6\times10^{18}$\,cm$^{-2}$ is estimated; for the latter, \citet{kil07} 
estimated $N_{\rm H} \approx 2\times10^{18}$\,cm$^{-2}$.  It is noteworthy that column densities differing by a factor of two lead 
to only a few percent difference in the calculated limiting fluxes, and that the accretion column itself is not a significant source 
of self-absorption.

The fluxes of the model plasmas were calculated for emission temperatures of 1, 4, and 10\,keV.  These span the range 
of characteristic single temperatures seen in disk-accreting cataclysmic variables (see \citealt{bas05}).  Polars (magnetic
cataclysmic variables) with column accretion can have higher characteristic temperatures, but low-state magnetic systems tend 
to have temperatures at the lower end \citep{muk17}.  The characteristic temperature seen in X-rays will be lower than the shock 
temperature, and the true spectrum will be something like a cooling flow, with emission from a range of temperatures.  The X-ray
emission spectra are shown Figure \ref{fig2} and plot both solar and bulk Earth composition models for the full temperature range.
The models were then scaled to match the {\em XMM} upper bound count rate for each source, yielding upper limit detector and
bolometric fluxes for each model listed in Table \ref{tbl2}.  While there is a visible difference between the spectra of solar and bulk 
Earth composition plasmas, the resulting flux limits differences are less than 20\% in the $kT=1$ and 4\,keV cases, and within a 
factor of two for 10\,keV.

\begin{table*}
\begin{center}
\caption{{\em XMM} Data and {\sc xspec} Model X-Ray Flux Limits\label{tbl2}}
\begin{tabular}{@{}ccccccccccc@{}}

\hline
\hline

Energy 		&Source	&Bkgd	&Upper$^{\rm a}$	&Upper$^{\rm a}$	&\multicolumn{2}{c}{$F_{\rm lim}$ ($kT=1$\,keV)}
															&\multicolumn{2}{c}{$F_{\rm lim}$ ($kT=4$\,keV)}
															&\multicolumn{2}{c}{$F_{\rm lim}$ ($kT=10$\,keV)}\\
Range		&Region	&Region	&Bound			&Rate 			&In-Band	&Total	&In-Band	&Total	&In-Band	&Total\\
\cline{2-4}
\cline{6-11}

\rule{0pt}{3ex} 

(keV)		&\multicolumn{3}{c}{(counts)} 		&(counts\,s$^{-1}$)		&\multicolumn{6}{c}{($10^{-15}$\,erg\,cm$^{-2}$\,s$^{-1}$)}\\
\hline

			&\multicolumn{3}{c}{\bf 1145+017 (126.0\,ks)}				&&\multicolumn{6}{c}{\bf Solar composition}\\
0.3-2.0		&64		&64			&15		&$1.2\times10^{-4}$		&0.4		&0.7		&0.5		&1.2		&0.5		&2.1\\
0.3-10.0		&126		&134			&17		&$1.3\times10^{-4}$		&0.5		&0.7		&0.9		&1.1		&1.1		&1.6\smallskip\\

															&&&&&\multicolumn{6}{c}{\bf Bulk Earth composition}\\
			&		&			&		&					&0.4		&0.6		&0.4		&1.4		&0.4		&3.7\\
			&		&			&		&					&0.5		&0.6		&1.0		&1.2		&1.7		&2.4\\

\hline

			&\multicolumn{3}{c}{\bf 1729+371 (20.4\,ks)}				&&\multicolumn{6}{c}{\bf Solar composition}\\
0.3-2.0		&7		&10			&5		&$2.3\times10^{-4}$		&0.9		&1.3		&0.9		&2.4		&0.9		&4.0\\
0.3-10.0		&16		&19			&7		&$3.2\times10^{-4}$		&1.3		&1.7		&2.1		&2.5		&2.6		&3.8\smallskip\\

															&&&&&\multicolumn{6}{c}{\bf Bulk Earth composition}\\
			&		&			&		&					&0.8		&1.1		&0.8		&2.7		&0.8		&7.2\\
			&		&			&		&					&1.2		&1.5		&2.5		&2.9		&4.2		&5.8\\
			
\hline

			&\multicolumn{3}{c}{\bf 2326+049 (22.5\,ks)}				&&\multicolumn{6}{c}{\bf Solar composition}\\
0.3-2.0		&16		&6			&18		&$8.0\times10^{-4}$		&3.6		&5.4		&3.9		&10		&3.9		&17\\
0.3-10.0		&25		&11			&23		&$1.0\times10^{-3}$		&4.8		&6.6		&8.0		&9.7		&10		&14\smallskip\\

															&&&&&\multicolumn{6}{c}{\bf Bulk Earth composition}\\
			&		&			&		&					&3.4		&4.7		&3.3		&11		&3.6		&30\\
			&		&			&		&					&4.4		&5.9		&9.4		&11		&16		&22\\

\hline
\hline

\end{tabular}
\end{center}

\flushleft
$^{\rm a}$ These are 90\% confidence bounds, calculated following \citet{kra91}, for the combined MOS1 and MOS2 detectors.

\smallskip
{\em Note.}  The in-band flux calculations include absorption along the line of sight due to interstellar hydrogen, while the 
bolometric (total) fluxes do not.

\end{table*}

\subsection{{\em XMM} Optical Monitoring Data}

\begin{figure}
\includegraphics[width=84mm]{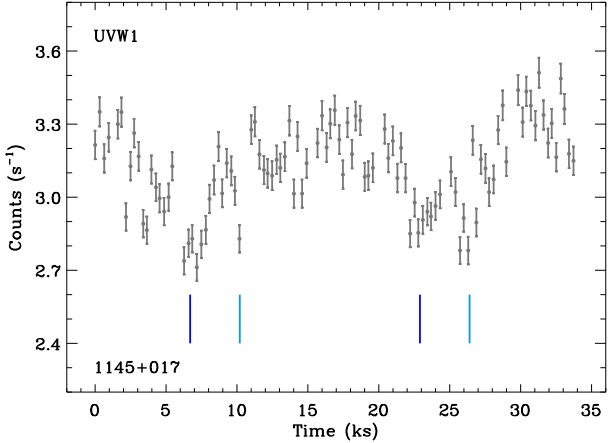}
\caption{Optical monitoring data from {\em XMM} covering the first 35\,ks of the X-ray observations for 1145+017 using the 
$UVW1$ filter.  The raw data were sampled at 10\,s, but have been re-binned in the plot to 300\,s to reduce noise and reveal 
structure in the crude light curve.  There are two broad sets of depressions in counts, with potentially two sub-structures each 
marked in dark and light blue, where these pairs are roughly separated by 4.5\,h = 16.2\,ks.  It is difficult to establish a baseline 
flux from these data, but the peak-to-trough distance is 23\% of the peak value, consistent with dimming activity throughout the 
X-ray exposure.}
\label{fig3}
\end{figure}

Data from the Optical Monitor (OM) telescope was also collected for 1145+017.  Observations were taken in photon-counting 
(FAST) mode with the {\em UVW1} filter, at an effective wavelength of 2910\,\AA.  Unfortunately, there was a telescope fault after 
the first 35\,ks and data were only recorded for the first $1/4$ of the total X-ray exposure.  Despite this, the OM data are plotted in 
Figure \ref{fig3} and show that the source was experiencing dimming events during this time.  The raw data were sampled at 10\,s 
cadence but had to be binned to 300\,s intervals for features to emerge, but demonstrate there were flux drops by at least 20\% 
and likely reveal two repeating structures. 

\section{Results and Context}

\subsection{Constraints on Weak Magnetism}

The observational incidence of magnetism among white dwarfs is well represented in the literature, but is fraught with detection 
and selection biases so that a coherent picture spanning cooling age and atmospheric composition has not yet emerged 
\citep{sch95,sch03,kaw07,hol15}.  While the origin and manifestation of magnetic fields in degenerates is beyond the scope 
of this paper, the subset of relatively weak, kG-order fields at polluted white dwarfs has implications for circumstellar disk 
structure and accretion onto the surface \citep{met12}.  Field strengths as small as 0.1--1\,kG can truncate a disk at the Alfv\'en 
radius or prevent its formation as in polars, and will result in accretion near freefall velocities onto magnetic polar regions, as 
opposed to equatorial accretion in a boundary layer at lower, sub-Keplerian velocities.

There are now a significant number of metal-enriched white dwarfs that exhibit (weak) magnetic fields, and this is potentially a 
detection bias as Zeeman splitting can be more readily detected in narrow features from multiple species of heavy elements,
than in stars with only weak Balmer lines, or no lines as in cool helium atmospheres \citep{kaw14,hol15}.  Nevertheless, if there
is a link between the debris observed to pollute white dwarf atmospheres, and the prevalence or strength of stellar magnetic 
fields, it would suggest that closely-orbiting planets or their engulfment during a common envelope can generate sustained 
(weak) magnetism in white dwarfs \citep{far11,kis15}.

This process would be a planetary-mass analog to strong magnetic field generation via mergers 
and common envelope evolution of stellar-mass companions \citep{tou08,nor11}.  The results discussed below are the part 
of a concerted effort to determine the frequency of weak magnetic fields among polluted white dwarfs.

For the measurements and upper limits obtained, a tilted and centered magnetic dipole will have a surface polar field strength
that is related to the maximum observed value of the surface-averaged field by the following inequality \citep{aur07}
\begin{equation}
B_* \gtrsim 3.3 \, \langle B_{\rm z} \rangle_{\rm max}
\label{eqn2}
\end{equation}
This is useful in the case where multi-epoch detections and stellar rotation rates are not available, as in this study.  Thus
a lower limit on the surface dipole component of the magnetic field can be estimated from circular spectropolarimetry.

{\bf 0322--019}.  This star was first shown to be weakly magnetic via Zeeman splitting in multiple metal and H$\alpha$
in an optical spectrum with resolving power $R\approx40\,000$ and S/N $\approx100$ \citep{far11}.  The nature of the 
weak splitting due to a $B_*\approx120$\,kG field only became apparent within a combined dataset consisting of two dozen 
individual, coadded spectra which totaled 6.0 hours of exposure.  The Ca\,{\sc ii} splitting had been previously detected in 
high-resolution data, but attributed to binarity \citep{zuc03}.  Figure \ref{fig1} plots the second data set of spectropolarimetry 
for this white dwarf, and shows that $P_{\rm v}$ is clearly non-zero in the second dataset.  Although the Zeeman splitting is 
unresolved in the FORS2 Stokes $I$ spectra, this magnetic field detection was achieved in just over 20\,min of on-source time 
(cf.\ the UVES detection).  $P_{\rm v}$ was determined only from the region of Ca\,{\sc ii} H, K, and Ca\,{\sc i} 4226\,\AA.  
The surface-averaged, longitudinal field detected is $16.5\pm2.3$\,kG, which is a factor of several smaller than the intrinsic 
field strength estimated from Zeeman splitting, and thus consistent with Equation \ref{eqn2}.  The first observation of this star 
did not result in a detection and is likely due to a chance alignment with the magnetic equator.

{\bf 1145+017}.  There are four relatively deep observations of this iconic source, but only in the third data set is there 
potentially real signal.  However, as discussed in \S2.1, anything below $5\upsigma$ cannot be viewed as a confident 
detection and hence the $3.5\upsigma$ result for the third epoch should be viewed as a promising result that requires 
confirmation with additional observations.  The third data set for this star is plotted in Figure \ref{fig1} and shows the
weak but non-zero slope in $P_{\rm v}$.  Adopting $\langle B_{\rm z} \rangle_{\rm max} \approx 1$\,kG, then the minimum 
dipolar field strength would be 3\,kG.

\begin{figure}
\includegraphics[width=84mm]{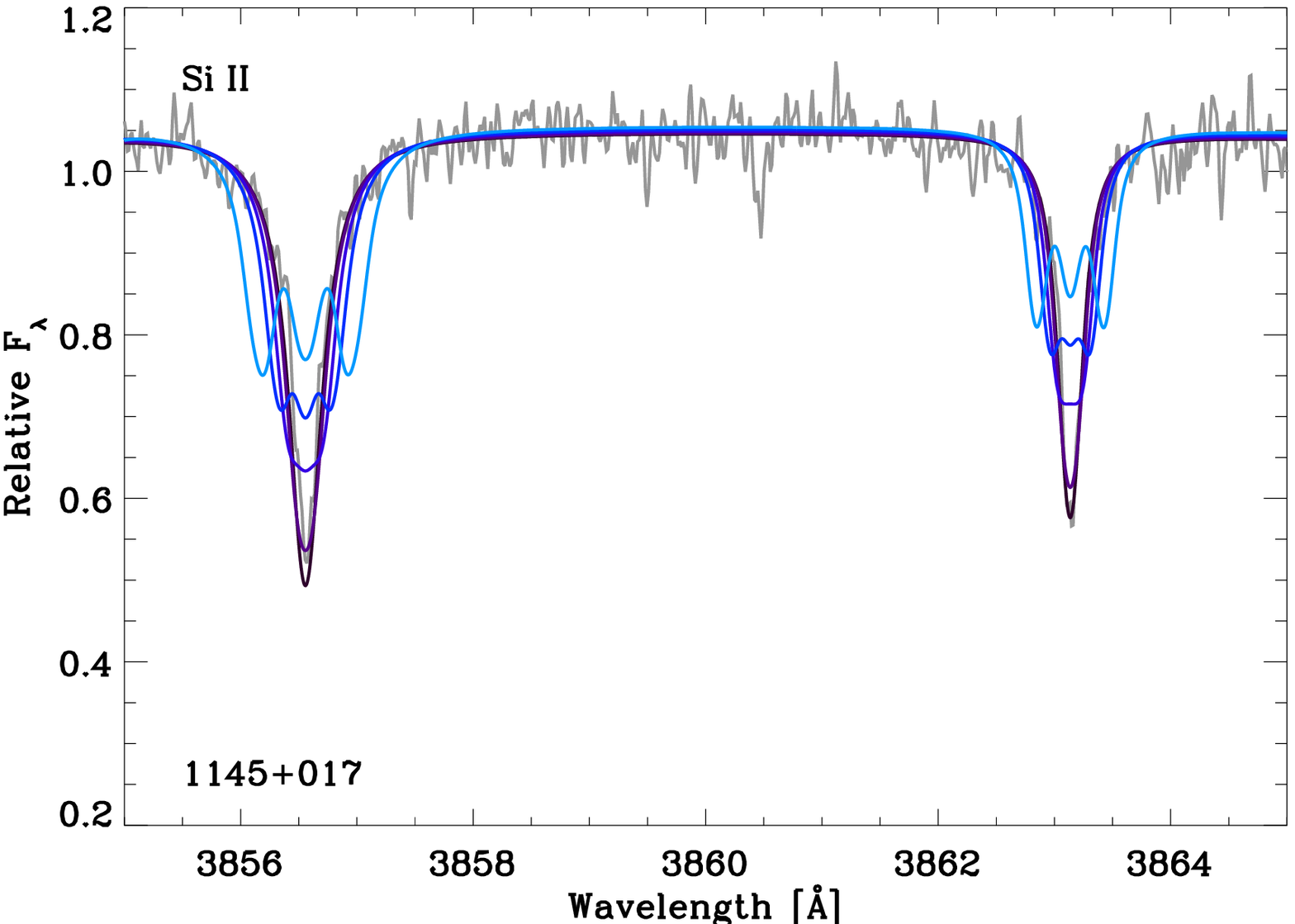}
\vskip 5pt
\includegraphics[width=84mm]{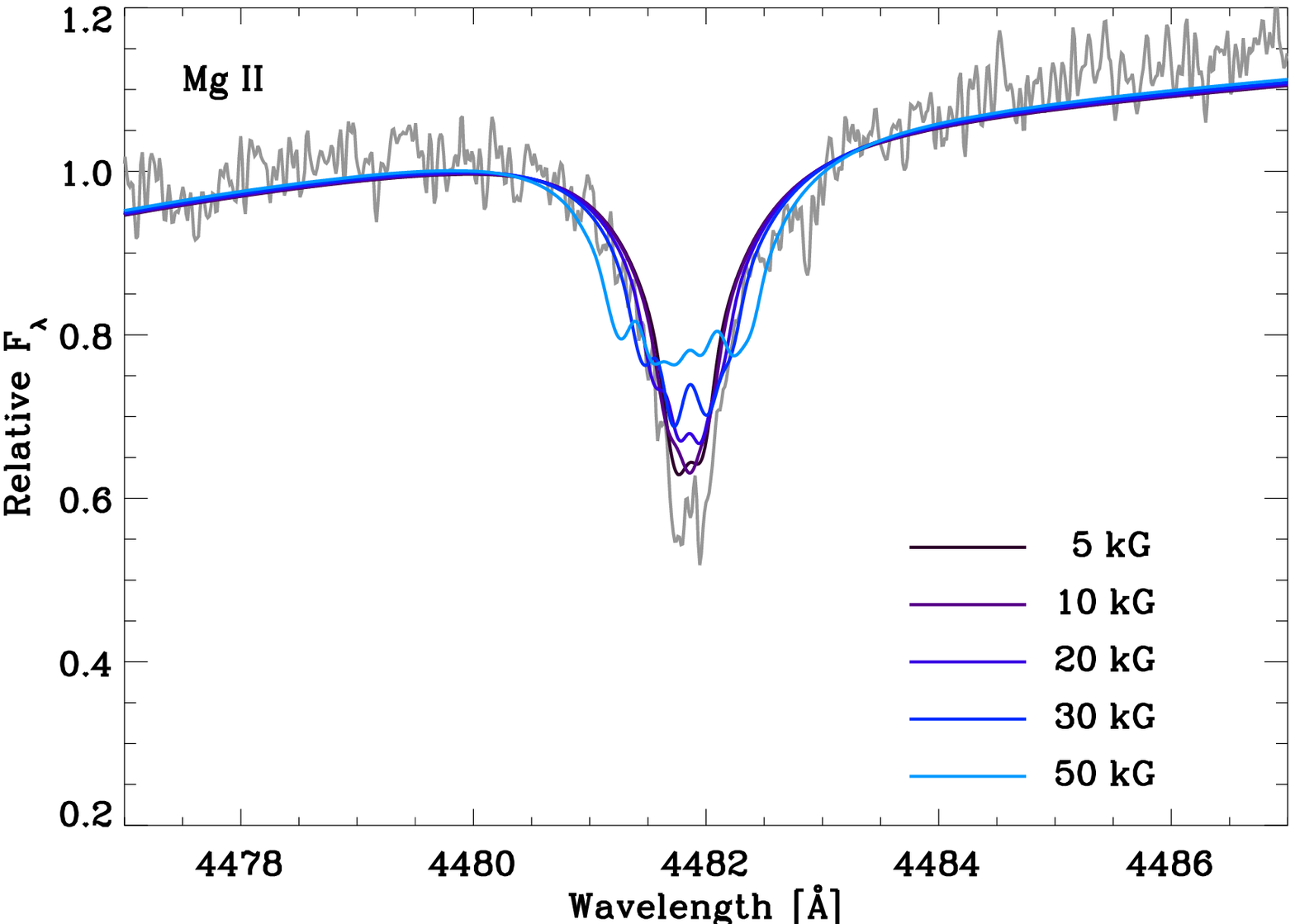}
\caption{Magnetic stellar atmosphere models used to estimate an upper limit to the mean magnetic field modulus for an 
intrinsic (dipolar) field for 1145+017.  Keck HIRES data with resolving power $R\approx40\,000$ \citep{xu16} are shown in 
grey for two regions containing the strongest metal absorption features that are purely photospheric, and where the lines 
are isolated from other (atmospheric and circumstellar) absorbers.  A synthetic stellar spectrum was modified for purely 
radial field strengths of $B_*= 5$, 10, 20, 30, and 50\,kG \citep{koc10} with the resulting spectra overplotted as the colors 
shown in the legend.  These model spectra indicate that the effects of Zeeman splitting should have been observed for field 
strengths above 20\,kG.
\label{fig4}}
\end{figure}

It is important to place the best possible constraint on the magnetic field of this star for the X-ray analysis that follows.  In 
order to complement the limits provided by circular spectropolarimetry, the published HIRES spectrum with resolving power 
$R\approx40\,000$ \citep{xu16} was examined for any indication of Zeeman splitting.  While none is evident, a series of model 
stellar atmospheres with increasing magnetic field strength was generated in order to place upper limits via the absence of 
Zeeman-split lines.  The {\sc llmodels} stellar atmosphere code \citep{shu04} was used, together with the published stellar 
parameters abundances.  From this baseline model atmosphere, synthetic Stokes $I$ stellar spectra were computed using 
the {\sc synmast} code \citep{koc10}, for purely radial (dipolar) magnetic fields of increasing strength.  The data and models 
are shown in Figure \ref{fig4} for two narrow wavelength regions with strong photospheric lines of Mg\,{\sc ii} and Si\,{\sc ii} 
that are well isolated from circumstellar, as well as additional stellar absorption.  The results of the magnetic modeling 
indicates that fields as large as 30\,kG would be obvious, and hence a more realistic upper limit is $B_* < 20$\,kG.

{\bf 1929+011}.  The highly polluted and $T_{\rm eff}\approx21\,000$\,K white dwarf has no previous observations using
circular spectropolarimetry.  The single observational limit here is not highly constraining, but taken at face value suggests
that a dipolar field on the order of tens of kG remains possible.  High-resolution UVES spectra of this star exist \citep{ven10}, 
which have nearly identical resolving power ($R\approx40\,000$) to the HIRES data analyzed above.  Modeling similar to 
that performed for 1145+017 was carried out for the UVES data for the strong Mg\,{\sc ii} 4482\,\AA \ feature (not shown), 
and a similar upper limit of 20\,kG is estimated.

{\bf 2105--820}.  This star was a magnetic suspect first identified during a search for rotational broadening in the NLTE cores 
of H$\alpha$, where it shows a clearly flattened core shape consistent with a $B_*= 43$\,kG intrinsic field \citep{koe98}. This 
has been confirmed with higher resolution data taken with UVES for the SPY survey \citep{koe09}, where it also exhibits a 
flattened core in the weak but clearly detected Ca\,{\sc ii} K absorption line \citep{koe05}, demonstrating that the metal is 
unambiguously photospheric.  

This white dwarf was previously detected in circular polarization over the higher Balmer lines, where five detections were 
obtained for surface-averaged fields in the range 8--11\,kG \citep{lan12}.  The data over the H$\alpha$ region are shown 
in Figure \ref{fig1} and are consistent with a longitudinal field of 5\,kG, which is roughly half that detected consistently over 
several days and longer by \citet{lan12} using FORS1.  This discrepancy is not actually meaningful, however, as studies 
have shown that even within a single instrument, each specific grating and wavelength setting define a specific instrumental 
system for measuring $\langle B_{\rm z} \rangle$ \citep{lan14}.  Owing to various differences induced by individual setups, 
measurements made with different systems are not directly comparable except for general magnitude.  Nevertheless, the 
field detected via H$\alpha$ spectropolarimetry leads to a lower limit on the dipolar magnetic field strength of roughly 17\,kG
(Equation \ref{eqn2}), which is consistent with the $B_*=43$\,kG estimated from Zeeman splitting.

{\bf 2326+049}.  This is one of the most well studied degenerate stars known and is the prototype dusty and polluted 
white dwarf \citep{zuc87,koe97}.  Circular spectropolarimetry was performed on this target over two decades prior, 
resulting in a non-detection with a $5\upsigma$ error of 64\,kG \citep{sch95}.  Table \ref{tbl1} indicates that the FORS2 
$5\upsigma$ upper limit is 2.5\,kG, consistent with both $\langle B_{\rm z} \rangle$ and $\langle N_{\rm z} \rangle$ and 
their dispersions.  As with the other stars in the sample, 2326+049 also has high-resolution optical spectra that can be 
used to place magnetic field limits via the absence of Zeeman splitting.  Using the same methodology as above for 
Ca\,{\sc ii} 3968\,\AA \ (not shown), an upper limit field strength of 20\,kG is estimated, consistent among all the stars 
with similarly high-resolution spectral data.

Although spectropolarimetry was not obtained for 1729+371, a similar estimate of $B_* < 20$\,kG is adopted, as this
polluted white dwarf also has $R\approx40\,000$ HIRES data with no evidence of Zeeman splitting \citep{zuc07}.

%
%


%
%

%
%

%
%

\subsection{Interpretation of X-ray Upper Limits}

\begin{figure}
\includegraphics[width=84mm]{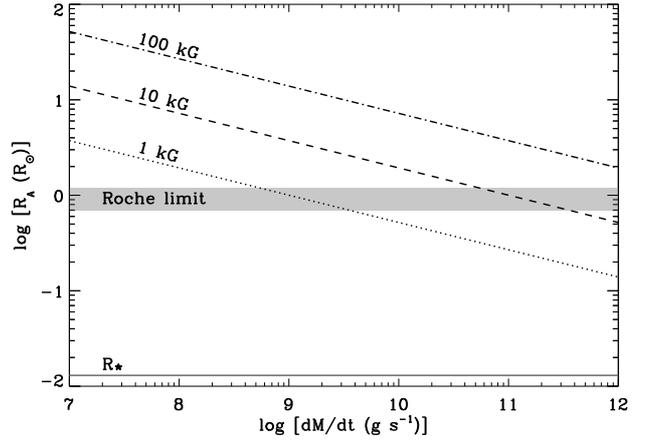}
\caption{Alfv\'en radius as a function of gas accretion rate $\mdot$ from Equation \ref{eqn3} for white dwarfs with fixed magnetic 
field strengths of $B_*=1$, 10, and 100\,kG.  Shown for comparison is a grey region whose height corresponds to the full range
of Roche radii for asteroids of mean density $\uprho = 1-5$\,g\,cm$^{-3}$.  Magnetic fields as weak 1\,kG or weaker will have a
profound effect on the behavior of gas interior to 1\,$R_{\odot}$ when $\mdot \lesssim 10^{10}$\,g\,s$^{-1}$, and especially as 
the material accretes within several $R_*$.}
\label{fig5}
\end{figure}

This section presents a theoretical framework in which to interpret the X-ray upper limits at polluted white dwarfs including 
1145+017.  The Alfv\'en radius for a gas disk accreting at a rate $\mdot = \mdot_{10}\times 10^{10}$\,g\,s$^{-1}$ is given by 
(e.g.\ \citealt{gho78})
\begin{eqnarray}
R_{\rm A} &=& \left(\frac{3B_{*}^{2} \, R_{*}^{6}}{2\mdot\sqrt{GM_{*}}}\right)^{2/7} \approx 0.52R_{\odot}\left(\frac{B_{*}}{\rm kG}\right)^{4/7}\mdot_{10}^{-2/7}
\label{eqn3}
\end{eqnarray}
where $B_{*}$ is the surface polar magnetic field strength of a white dwarf with assumed mass $M_{*} = 0.6\,M_{\odot}$ 
and radius $R_{*} = 9.0\times10^{8}$\,cm.  Figure \ref{fig5} plots $R_{\rm A}$ as a function of accretion rate for a range of
magnetic field strengths representative of detections and upper limit estimates for the observed sample.  Thus for the ongoing 
accretion rates thought to be characteristic of polluted white dwarfs, the Alfv\'en radius will reside near or exterior to the Roche 
radius \citep{met12}, and further outside of where the disk becomes dominated by (sublimated) gas.  Furthermore, $R_{\rm A}$ 
will remain well above the stellar surface unless the star is essentially non-magnetic and $B_{*} < 1$\,G, or the accretion rate
is extreme at $\mdot > 10^{16}$\,g\,s$^{-1}$.

For typical (single) white dwarfs, rotation periods $2\uppi/\Omega_{*}$ are many hours to days \citep{her17}, and hence
the co-rotation radius $R_{\rm c} = (GM_{*}/\Omega_{*}^{2})^{1/3}$ will also be outside the expected radius of the gas disk.
The accretion flow should therefore divert onto the stellar magnetosphere near the radius $R_{\rm A}$ (without a propellor), 
interior to which matter will be placed onto field lines leaving the magnetic polar region at a characteristic latitude $\uptheta_{\rm m} 
\approx \sin^{-1}(\sqrt{R_{*}/R_{\rm A}}) \approx \sqrt{R_{*}/R_{\rm A}}$.  The accretion column will therefore cover a fraction 
of the stellar surface crudely given by
\begin{equation}
f_{\rm m} = \frac{2 \uppi \, \uptheta_{\rm m}^{2}}{4\uppi} \approx \frac{1}{2}\frac{R_{*}}{R_{\rm A}} \approx 0.013\left(\frac{B_{*}}{\rm kG}\right)^{-4/7}\mdot_{10}^{2/7}
\label{eqn4}
\end{equation}
This expression assumes a magnetic axis aligned with the disk angular momentum, where a large misalignment will decrease 
$f_{\rm m}$ by factors of order unity.  Furthermore, observations of X-ray emitting spots on white dwarfs demonstrate that the 
accretion geometry can be substantially more complex than this (see \citealt{muk17}, and references therein).  Nevertheless, 
Equation \ref{eqn4} serves as a useful first approximation for the fraction of the stellar surface subject to accretion infall.

Matter falling onto the polar cap will have free-fall velocity $v_{\rm ff} \approx \sqrt{2GM_{*}/R_{\rm sh}} \approx \sqrt{GM_{*}
/R_{*}} \approx 3000$ km s$^{-1}$, where the shock radius $R_{\rm sh}$ is taken to be twice the stellar radius.  Given the 
range of relevant accretion rates, the shock will be adiabatic and thus not immediately collapse into a thin radiative shock 
close to the stellar surface.  The infalling gas will be shock-heated to a temperature
\begin{equation}
kT_{\rm sh} \approx \frac{3}{16} \, \upmu m_{\rm p} v_{\rm ff}^{2} \approx \frac{3}{8}\frac{GM_{*}m_{\rm p}}{R_{*}} \approx 37\,{\rm keV}
\label{eqn5}
\end{equation}
where $\upmu \approx 2$ is the mean molecular weight for fully-ionized matter of bulk Earth composition.  The post-shock 
gas will be compressed to a density of
\begin{equation}
\uprho_{\rm sh} \approx \frac{\mdot}{\uppi \, R_{\rm sh}^{2} \, f_{\rm m} \, v_{\rm ff}} \approx 3.6\times 10^{-16}\mdot_{10} \, \left(\frac{f_{\rm m}}{0.01} \right)^{-1} {\rm g\,cm^{-3}}
\label{eqn6}
\end{equation}
Gas will cool behind the shock radiatively at approximately constant pressure $P \propto \uprho T$.  Thus, neglecting 
recombination effects, as the gas cools to temperature $T$ from its initial $T_{\rm sh}$ (Equation \ref{eqn5}), it will compress 
according to
\begin{equation}
T\uprho = T_{\rm sh}\uprho_{\rm sh}
\label{eqn7}
\end{equation}

In order to settle onto the stellar surface, the gas must release a total radiative luminosity of
\begin{equation}
L_{\rm tot} = \frac{GM\mdot}{R_{*}} \approx 8.9\times 10^{26}\mdot_{10}\,{\rm erg\,s^{-1}}
\label{eqn8}
\end{equation}
Three gas cooling mechanisms within the accretion column will determine the wavelength regimes for this emission: (1) 
bremsstrahlung (free-free emission); (2) atomic (i.e.\ line) emission; and (3) cyclotron emission.  The cooling rate per unit volume 
is commonly written as $\qdot = -n^{2}\upLambda,$ where $n \equiv \uprho/m_{\rm p}$ and $\upLambda$ is the cooling function.  
Cooling rates in general will differ from that of solar metallicity gas due to the distinct composition of the accreting (planetary) 
matter, and here is taken as bulk Earth \citep{mcd00}, as is observed to dominate the heavy element mass in more than one 
dozen white dwarfs with detailed measurements \citep{jur14}.

\begin{figure}
\includegraphics[width=84mm]{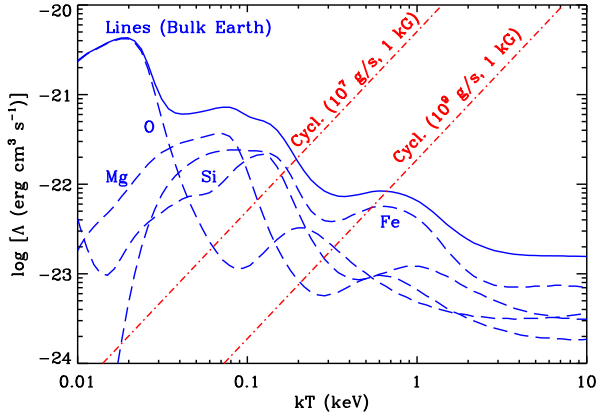}
\caption{Cooling function $\Lambda$ for different processes in the accretion column as a function of the gas temperature. Atomic 
cooling rates for fully-ionized matter with bulk Earth composition \citep{mcd00} are shown with blue lines; dashed curves show 
the contributions of individual elements and a solid line shows the total cooling rate.  Dot-dashed red curves show the cyclotron 
cooling rate for different assumptions about accretion rate $\dot M = 10^{7}$ and $10^{9}$\,g\,s$^{-1}$, for an assumed surface 
magnetic field strength of $B_{*} = 1$\,kG.}
\label{fig6}
\end{figure}

At low temperatures and high density, atomic cooling will predominate.  Figure \ref{fig6} shows the atomic cooling rate for bulk 
Earth composition, calculated assuming collisional ionization equilibrium and fully ionized gas (a reasonable approximation at 
the high temperatures of interest) from \citet{sch09}.  Owing to the large number of atomic transitions, cooling from Fe will 
determine the total cooling rate across all temperatures of interest, which also greatly exceeds the free-free cooling rate.

However, at low densities and high temperatures, cyclotron cooling can dominate over atomic cooling if the magnetic field 
is sufficiently strong.  The cyclotron cooling rate behind the shock near the stellar surface is approximately given by
\begin{eqnarray}
\upLambda_{\rm B} &\approx& \frac{3}{2}\frac{\upsigma_{\rm T}}{c}\frac{B_{\rm sh}^{2}}{8\uppi}\frac{m_{\rm p}}{m_{\rm e}}\left(\frac{ kT}{\uprho}\right) = \frac{3B_{*}^{2}}{128\uppi}\frac{\upsigma_{\rm T}}{c}\frac{m_{\rm p}}{m_{\rm e}}\left(\frac{kT_{\rm sh}}{\uprho_{\rm sh}}\right)\left(\frac{T}{T_{\rm sh}}\right)^{2} \nonumber \\
&\approx& 5\times 10^{-20}\mdot_{10}^{-1} \left(\frac{f_{\rm m}}{0.01} \right) \left(\frac{B_{*}}{\rm kG}\right)^{2}\left(\frac{T}{T_{\rm sh}}\right)^{2}\,{\rm erg\,cm^{3}\,s^{-1}}  \nonumber \\
&\approx& 6\times 10^{-20} \, \mdot_{10}^{-5/7}\left(\frac{B_{*}}{\rm kG}\right)^{10/7}\left(\frac{T}{T_{\rm sh}}\right)^{2}\,{\rm erg\,cm^{3}\,s^{-1}}
\label{eqn9}
\end{eqnarray}
where $\upsigma_{\rm T}$ is the Thomson cross section.  In the second and final lines Equations \ref{eqn6} \& \ref{eqn7}
have been used, and in the final line $f_{\rm m}$ has been substituted from Equation \ref{eqn4}.  The magnetic field strength 
$B_{\rm sh}$ behind the shock at $r \approx 2R_{*}$, is assumed to be 8 times lower than the surface value $B_{*}$ due to 
the $\propto 1/r^{3}$ dilution for a dipole field.

Fitting the atomic cooling rate to a power law of $\Lambda = 1.0\times 10^{-23}(kT/{\rm 10\,keV})^{-0.8}\,{\rm erg\,cm^{3}\,s^{-1}}$,
versus the comparatively sensitive temperature dependence of $\upLambda_{\rm B} \propto T^{2}$, we estimate that cyclotron cooling 
will dominate above a critical temperature
\begin{equation}
T_{\rm B} \approx 1.1\,\mdot_{10}^{0.26}\left(\frac{B_{*}}{\rm kG}\right)^{-0.51}{\rm keV},
\label{eqn10}
\end{equation}
In cases when $T_{\rm sh} \gtrsim T_{\rm B}$ it is expected that only a fraction $\sim T_{\rm B}/T_{\rm sh}$ will emerge as 
line emission at temperatures $T \lesssim T_{\rm B}$.  Translating to the regime of interest for the {\em XMM} observations,
the total X-ray luminosity (when $T_{\rm B} \lesssim T_{\rm sh} \approx 37$ keV) will be given by
\begin{equation}
L_{\rm X} = \frac{T_{\rm B}}{T_{\rm sh}}L_{\rm tot} \approx 3\times 10^{25}\mdot_{10}^{1.26}\left(\frac{B_{*}}{\rm kG}\right)^{-0.51}\,{\rm erg\,s^{-1}}
\label{eqn11}
\end{equation}
The remaining luminosity will emerge as cyclotron emission, at much lower radio frequencies of
\begin{equation}
\upnu \gtrsim \frac{eB_{*}}{2\uppi m_{\rm e} c} \approx 3\,\left(\frac{B_{*}}{\rm kG}\right)\,{\rm GHz}
\label{eqn12}
\end{equation}
Any cyclotron radiation has a maximum flux density set by the blackbody equivalent at $T_{\rm sh}$.  As flux scales linearly 
with temperature in the Rayleigh-Jeans regime, the most optimistic cyclotron flux can only be $T_{\rm sh}/T_{\rm eff}\sim10^4$ 
times brighter than the stellar photosphere.  This puts optimistic radio fluxes in the nJy range for typical white dwarfs and thus 
not detectable with current radio facilities.

\subsection{Upper Limit Accretion Rates}

Upper limit accretion rates were calculated for the three X-ray observed white dwarfs in two regimes: where weak magnetic 
fields of order 1\,kG may play a role Equation \ref{eqn11} was used, and a non-magnetic calculation was done using Equation 
\ref{eqn8}.  As the temperature of the emission is empirically unconstrained, both X-ray luminosity and accretion rate limits 
were derived for the full range of $kT$ considered here.  Calculations were based on the model spectra for material with bulk 
Earth composition, and the most stringent bolometric X-ray flux constraint for each $kT$ value was taken as input (e.g.\ the 
0.3--2.0\,keV energy range limits for $kT=1$\,keV).  Upper bound, total X-ray fluxes were transformed into luminosities using 
the best available distance estimate for 1145+017 (174\,pc; \citealt{van15}), and the most recent parallaxes for 1729+371 
(50.6\,pc; \citealt{kil08}) and 2326+049 (17.6\,pc; \citealt{sub17}).  The calculated upper limit X-ray luminosities and mass 
accretion rates are given in Table \ref{tbl3}, with $L_{\rm X}$ values listed as totals, whereas $\dot M$ limits account for
half the luminosity being directed away from the observer.

{\bf 1145+017}.  Unless this star has a magnetic field strength that is significantly greater than 1\,kG, then the accretion rate 
based on the models presented here cannot be as high as 10$^{12}$\,g\,s$^{-1}$.  Furthermore, if the white dwarf is essentially 
non-magnetic, then the current accretion rate should be less than 10$^{11}$\,g\,s$^{-1}$ and thus consistent with the rate inferred 
from the atmospheric metal abundances under the assumption of a steady-state balance between accretion and diffusion.  
However, 1145+017 has an atmosphere dominated by helium, where a typical heavy element persists for several $10^5$\,yr 
\citep{koe09} before fully sinking below the outer layers, and thus a steady-state is far from certain.  The current accretion rate 
could be significantly lower than the upper limits broadly set by the X-ray data and modeling, and even lower than more modest 
rates of order 10$^{10}$\,g\,s$^{-1}$ for a steady-state accretion regime.  

\begin{table}
\begin{center}
\caption{Benchmark X-Ray Luminosity and Accretion Rate Limits\label{tbl3}}
\begin{tabular}{@{}cccc@{}}

\hline
\hline

$kT$ 		&$L_{\rm X}$			&$\dot M$ ($B_*=1$\,kG)$^{\rm a}$ 	&$\dot M$ ($B_*=0$)$^{\rm b}$\\

\cline{3-4}
\rule{0pt}{3ex} 

 (keV)		&(10$^{27}$\,erg\,s$^{-1}$)	&\multicolumn{2}{c}{(10$^{10}$\,g\,s$^{-1}$)}\\

\hline

%
%
%
%
%

&\multicolumn{3}{c}{\bf 1145+017 (174\,pc)}\\

1		&2.1		&50		&4.7\\
4		&4.3		&89		&9.7\\ 
10		&8.7		&160		&20\\

\hline

&\multicolumn{3}{c}{\bf 1729+371 (50.6\,pc)}\\

1		&0.3		&12		&0.8\\
4		&0.8		&24		&1.9\\
10		&1.8		&44		&4.0\\

\hline

&\multicolumn{3}{c}{\bf 2326+049 (17.6\,pc)}\\

1		&0.2		&7.0		&0.4\\
4		&0.4		&14		&0.9\\
10		&0.8		&24		&1.9\\

\hline
\hline

\end{tabular}
\end{center}

\flushleft

\smallskip
$^{\rm a}$ Using Equation \ref{eqn11}.\\

$^{\rm b}$ Using Equation \ref{eqn8}.\\

{\em Note.}  All values are for accreting material of bulk Earth composition.  $L_{\rm X}$ is the bolometric luminosity due to
accretion, while $\dot M$ accounts for the fact that half of $L_{\rm X}$ is directed back into the star.

\end{table}

These findings are also consistent with an ongoing accretion rate below all the above estimates, and potentially on the order 
of 10$^{8}-10^{9}$\,g\,s$^{-1}$, which represent the upper end of rates confidently inferred to be ongoing for white dwarfs with 
infrared excess \citep{ber14}.  Nevertheless, the current mass of atmospheric metals divided by their sinking timescale does 
give a historical average rate of accretion over the past diffusion timescale, and in the case of 1145+017 suggests the system 
experienced some higher-rate episode(s) within the past Myr \citep{gir12}, where these are likely short-lived and stochastic 
events \citep{far12,wya14}.  Overall, these results support the possibility that 1145+017 is a relatively ordinary dusty and 
polluted white dwarf, but with a particular viewing geometry that reveals a spectacular light curve and absorption spectrum.

{\bf 1729+371 and 2326+049}.  The results presented here for these two stars can be compared directly with a previously 
published analysis of their {\em XMM} data, where \citet{jur09} find upper limit accretion rates of $2\times10^{10}$\,g\,s$^{-1}$ 
and $2\times10^{9}$\,g\,s$^{-1}$, respectively.  There are several noteworthy differences between the approach taken by that 
study, and the data and modeling done here.  First, upper bound count rates were converted to fluxes using {\sc PIMMS} as 
opposed to the more sophisticated {\sc XSPEC} analysis done here, where the former is restricted to solar ratios of the 
elements.  Second, any interstellar absorption was ignored (however, this simplification is probably fine as both stars are
within the Local Bubble), but is accounted for in Table \ref{tbl3}.  Third, in the previous study the bolometric corrections to 
the observed flux limits were made by assuming the accretion luminosity emerges solely as high-energy radiation (i.e.\ no 
stellar magnetism or cyclotron radiation), with half of the photons directed toward an observer, and half of these within the 
{\em XMM} bandpasses.  Here {\sc XSPEC} was used to construct a bolometric flux $F$ for each star and model, and 
these have been converted to luminosity using $L = F \times 4\pi d^2$, where $d$ is the distance to each star.

Given these potentially significant differences, it is perhaps surprising that the accretion rate limits reported by \citet{jur09} 
are comparable to the values listed in the fourth column of Table \ref{tbl3}.  For 1729+371, the upper bound counts of both 
studies are identical, and the non-magnetic limits calculated here are a match at 4\,keV.  Both sets of results are consistent 
with an ongoing accretion rate comparable to warm DAZ stars -- where a steady state is likely, and accretion rates can be 
inferred with confidence from metal abundances -- and no more than a few times $10^{9}$\,g\,s$^{-1}$ \citep{far12}.

However, for 2326+049 the non-magnetic limits on accretion rate found here for 4\,keV are roughly a factor of three larger
than those reported by \citet{jur09}, reflecting the fact that their upper bound count rates are a factor of three smaller than 
those listed in Table \ref{tbl2}.  Consistent with this, their limiting fluxes for this source are a factor of a few smaller than the 
in-band flux values listed in Table \ref{tbl2}.  As mentioned in \S2.2, no attempt was made here to remove or model the 
nearby X-ray source whose flux contaminated the aperture used to derive the upper bound counts here.

\begin{figure}
\includegraphics[width=84mm]{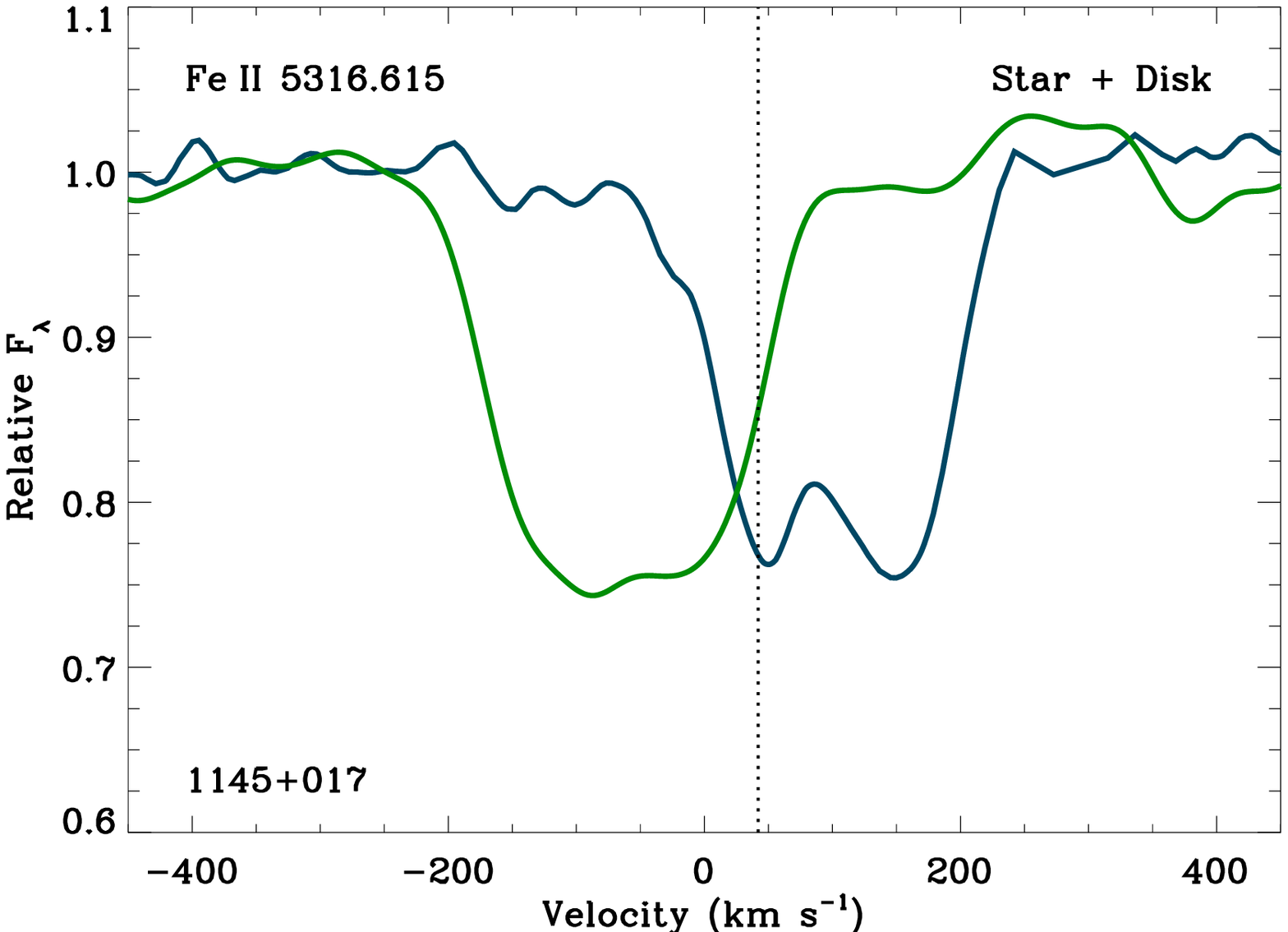}
\vskip 5pt
\includegraphics[width=84mm]{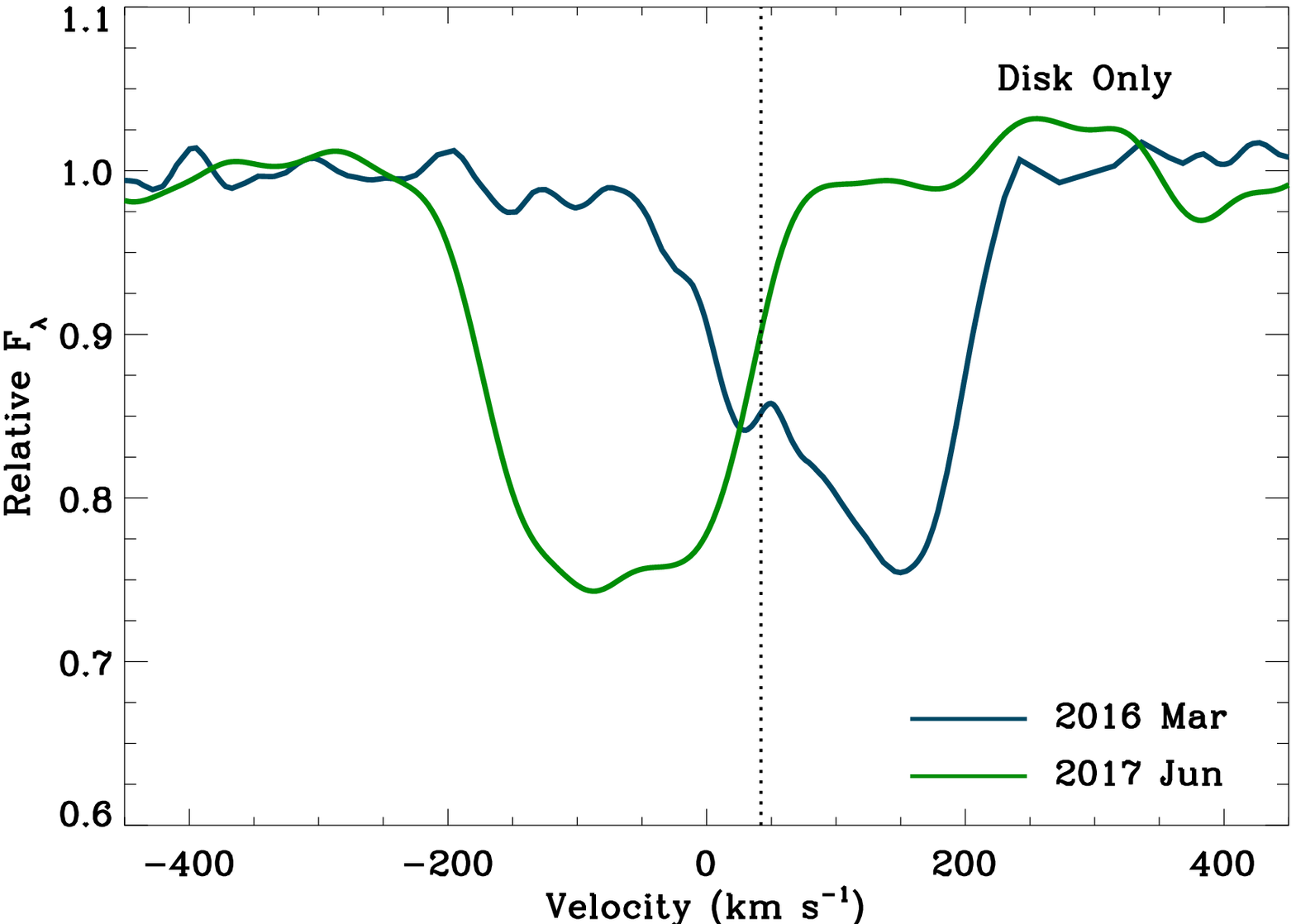}
\caption{Two epochs of spectroscopy for 1145+017 in the vicinity of the Fe\,{\sc ii} 5316\,\AA \ line, separated by 1.3\,yr.  VLT 
X-shooter data taken in 2016 March are plotted in dark blue, and Keck HIRES spectra taken in 2017 June are shown in green 
and have been re-binned to match the resolution of the X-shooter data.  The top panel shows the combined and normalized 
spectra as obtained, while the stellar atmosphere has been removed from the data shown in the bottom panel \citep{red17}, 
and the dotted lines denote the photospheric velocity.  The clear variation in disk geometry indicates that the red-shifted gas 
seen in 2015 and 2016 is not in the process of accreting onto the star at the observed velocities, and hence the high accretion 
rate calculated for this star based on the circumstellar gas mass and velocity \citep{xu16} is unlikely.
\label{fig7}}
\end{figure}

If the factor of a few difference between the studies is taken at face value, and is an accurate reflection of the upper bounds
to the counts from 2326+049 (based on undocumented but accurate corrections made by \citealt{jur09}), then that would fold 
into the upper limit accretion rates derived here.  A factor of three fewer in counts would translate into commensurately lower
X-ray luminosities and accretion rate limits.  \citet{xu14b} derive a steady-state accretion rate of $6.5\times10^{8}$\,g\,s$^{-1}$ 
for 2326+049 based on the detection of eight heavy elements, while the non-magnetic Table \ref{tbl3} values made smaller by 
a factor of three would yield a range of $1-6\times10^{9}$\,g\,s$^{-1}$.  While these values are all potentially consistent, it 
raises the possibility that accretion luminosity from 2326+049 may be detectable with a deep X-ray pointing.

\subsection{Additional Constraints on Accretion in 1145+017}

As part of an ongoing program to monitor the circumstellar absorption features seen in optical spectra of 1145+017, Keck / 
HIRES data were taken on 2017 Jun 27.  The data were taken and reduced in a manner identical to that described in detail 
in \citet{red17}.  A portion of these recent HIRES data are displayed in Figure \ref{fig7}, in the region surrounding the strong
circumstellar absorption feature from Fe\,{\sc ii} 5316\AA, and plotted alongside similar data taken 15 months prior.  As can 
be seen from these two sets of spectra, the distribution of velocities of the circumstellar gas has changed dramatically, and
the gas does not appear to be infalling.

\citet{red17} showed the circumstellar gas disk orbiting 1145+017 had an inner edge in the range $60-80\,R_*$ that was 
nearly constant from 2015 November to 2016 April, and one possible explanation was disk truncation due to a magnetic 
field.  While this possibility is consistent with the range of allowed $R_{\rm A}$ (see Figure \ref{fig5}) and the results found
here using spectropolarimetry, \citet{red17} concluded that the magnetospheric accretion model could not account for the 
particular shape of the observed circumstellar gas line profiles as well as the necessary ring eccentricity. 

The {\em XMM} observations of 1145+017 were designed to test the magnetospheric truncation possibility, as this model
contains analytical relationships between the observed gas velocity distribution via absorption, accretion rate and magnetic 
field at a given orbital radius.  This model predicted a range of accretion rates for 1145+017 in the range $2\times10^{11}-
5\times10^{12}$\,g\,s$^{-1}$, which are clearly ruled out by the X-ray data.  The new epoch of HIRES data also supports 
a gas disk behavior that is not strongly influenced by a magnetosphere via truncation at the inner edge, and which is also 
consistent with the upper limit magnetic field estimates from both the absence of Zeeman splitting and spectropolarimetry.

\section{Discussion and Conclusions}

\subsection{Accretion Overview}

There are currently few empirical constraints on the exact behavior of gas and dust -- production, evolution, and eventual 
accretion -- within planetary debris disks orbiting white dwarfs.  Specifically there are no direct measurements of instantaneous
accretion rate, only inferences based on the mass of atmospheric metals in stars where a steady-state is likely due to sinking 
timescales $t \ll 100$\,yr.  While these inferred, ongoing rates are limited to rates on the order of $10^9$\,g\,s$^{-1}$ and 
lower, there is compelling evidence for historical rates up to several orders of magnitude higher \citep{far12}, based on the
time-averaged accretion rates for those stars with sinking timescales up to order Myr \citep{gir12}.  Any high-rate episodes
must necessarily be short-lived to be consistent with the lack of detection among hundreds of stars where a steady-state
regime is favored \citep{koe05,gir11}.

Models for accretion onto polluted white dwarfs have demonstrated that Poynting-Robertson (PR) drag provides the rate
bottleneck for a disk dominated by particles \citep{raf11}.  Moreover, the infall rates predicted by modeling of {\em optically 
thick} disks are in excellent agreement with those inferred for systems likely to be accreting in a steady state \citep{far16}.
In contrast, the accretion rate predictions for {\em optically thin} disks are at best a factor of $10^2$ lower than that inferred
from observational data \citep{boc11}.  However, if an optically thin cloud or shell has significant vertical extent where all 
particles are unobscured and directly illuminated by starlight, then $\mdot = \tau L_* / c^2$ where $\tau$ is the fraction of 
intercepted starlight \citep{met12}, and this formulation appears consistent with the vertical extent estimated for the
circumbinary material polluting SDSS\,J155720.77+091624.6 \citep{far17a}.

The presence and generation of gas within a disk can strongly influence the evolution and accretion rate, especially in 
the regimes of a completely gaseous disk, where the viscous spreading timescale can be orders of magnitude shorter 
than that for PR drag on dust \citep{jur08}, or one where particles are strongly coupled to the gas \citep{met12}.  In such
cases, it has been speculated that accretion rates can soar to $10^{11}$\,g\,s$^{-1}$ or possibly even higher \citep{bea11}.
Although a small sample of three systems, the neither X-ray data or steady-state rates inferred from metal abundances have
yet to detect such high-rate accretion.

Recently, \citet{ken17} have modeled a narrow and mildly eccentric annulus containing a swarm of debris particles that
undergo a collisional cascade.  In contrast to other disk models, the collisional cascade produces disks with large vertical 
heights that are up to several times the stellar radius, and accretion rates that can exceed $10^{11}$\,g\,s$^{-1}$ by up to
two orders of magnitude, and persist for timescales up to $10^4$\,yr.  This collisional disk model thus requires replenishment 
on timescales where the cascade depletes the disk, and this could be accomplished with the stochastic infall of planetesimals 
as described by \citep{wya14}.

Overall, this study finds no evidence for accretion rates exceeding the $10^8-10^9$\,g\,s$^{-1}$ inferred by steady-state
calculations, but a larger sample of X-ray observations would provide better constraints on instantaneous accretion rates.

\subsection{Results Summary}

Using circular spectropolarimetry, robust detections of magnetic fields are found for 0322--019 and 2105--820, supporting 
earlier estimates from Zeeman splitting in dipole fields of $B_* \approx 120$ and 40\,kG respectively.  Neither of these 
polluted white dwarfs has an infrared excess, and while this is consistent with their Alfv\'en radii being larger than their 
Roche radii, it is not necessarily the case that the absence of infrared disk emission is due to magnetospheric disk truncation.  
In the case of 0322--019 with $T_{\rm eff}\approx5300$\,K, this star sits among a large class of metal-lined white dwarfs older 
than 0.5\,Gyr, where infrared excesses are rarely observed \citep{xu12,ber14}, and in which the heavy element sinking 
timescales are sufficiently long that accretion may have ended.  For 2105--820, it has a steady-state accretion rate of 
$3\times10^7$\,g\,s$^{-1}$ based on its calcium abundance, and likely below the ability of space- and ground-based
detection of its disk \citep{roc15,bon17}.

While there is a tentative indication of a spectropolarimetric signal from 1145+017, additional confirmation is necessary.
The high-resolution spectral modeling of several metal lines where Zeeman splitting is absent suggests the field cannot be
larger than around 20\,kG, and similarly for 1929+011 and 2326+049 based on comparable analyses of archival spectra.
Thus it appears the magnetically-trapped dust model of \citep{far17b} is likely ruled out in the case of the observed dimming
events towards 1145+017.  Further searches for weak magnetic fields in polluted white dwarfs can assess the applicability
of this model in other systems.

The study aimed to probe the role of magnetic fields in either the evolution of dust or its eventual accretion onto the surface 
of white dwarfs such as 1145+017, and to directly detect any accretion luminosity.  In the case of 1145+017, several favorable 
estimates have been given in the literature ranging from $10^{10}$\,g\,s$^{-1}$ to $10^{12}$\,g\,s$^{-1}$ \citep{van15,xu16,
gan16,rap16}.  As there is no confident indication of magnetic fields in the X-ray observed sample, upper limit accretion rates
are taken in the non-magnetic regime and appear to support none of the above estimates.  Rather, as discussed above, the
spectacular nature of 1145+017 appears to be its geometry, not its accretion rate.  If correct, the underlying parent body or
bodies inferred to be orbiting near 4.5\,h may be undergoing {\em collisions} rather than disintegration, as the latter implies
the system is observed at a special time, which may not be the case.

For 2326+049 the upper limit $\mdot$ values are consistent with that derived by \citet{xu14} for steady-state accretion, or
several times $10^{8}$\,g\,s$^{-1}$.  This is tantalizingly close to the most favorable X-ray limit if the upper bound count rate 
reported by \citet{jur09} is more accurate than that estimated here.  It is noteworthy that such a rate is consistent with 
predictions for optically thick disk models, and also the case where an optically thin disk has a significant vertical height 
and intercepts several percent of incident starlight.  However, in this case micron-sized disk particles should have been 
consumed by PR drag since its discovery \citep{zuc87}, unless they are being replenished.

\section*{Acknowledgements}

J. Farihi acknowledges funding from the STFC via an Ernest Rutherford Fellowship, as well as Columbia and Wesleyan 
Universities for visiting support during the preparation of the manuscript.  P. Wheatley was supported by STFC Consolidated 
grant ST/P000495/1.  S. Redfield, P. W. Cauley, and S. Bachman acknowledge the support from the National Science 
Foundation from grants AAG/AST-1313268 and REU/AST-1559865.  N. Achilleos was partly supported by STFC grant 
ST/N000722/1.  N. C. Stone received financial support from NASA through Einstein Postdoctoral Fellowship Award Number 
PF5-160145.

\bsp    
\label{lastpage}
\end{document}